\documentclass{svproc}

\usepackage{url}

\usepackage{graphicx}
\usepackage{amsmath}
\usepackage{amssymb}
\usepackage{booktabs}
\usepackage{array}
\usepackage{tikz}
\usetikzlibrary{shapes.geometric, arrows.meta, positioning, fit, backgrounds}

\begin{document}
\mainmatter

\title{An Agentic Multi-Agent Architecture for Cybersecurity Risk Management\thanks{Preprint. Submitted to AICTC 2026 (Springer LNCS).}}

\titlerunning{Agentic AI for Cybersecurity Risk Management}

\author{Ravish Gupta\inst{1} \and
Saket Kumar\inst{2} \and
Shreeya Sharma\inst{3} \and
Maulik Dang\inst{4} \and
Abhishek Aggarwal\inst{4}}

\authorrunning{R. Gupta et al.}

\institute{
Lead Software Engineer, BigCommerce; IEEE Senior Member\\
\email{ravishgupta@ieee.org}
\and
University at Buffalo, The State University of New York, Buffalo, NY, USA\\
\email{saketkmr.dev@gmail.com}
\and
Software Engineer, Microsoft\\
\email{shreeya2304@gmail.com}
\and
Senior Software Engineer, Amazon\\
\email{0111abhi@gmail.com, dangmaulik@gmail.com}
}

\maketitle

\begin{abstract}
Getting a real cybersecurity risk assessment for a small organization is
expensive---a NIST CSF-aligned engagement runs \$15,000 on the low end, takes
weeks, and depends on practitioners who are genuinely scarce. Most small
companies skip it entirely. We built a six-agent AI system where each agent
handles one analytical stage: profiling the organization, mapping assets,
analyzing threats, evaluating controls, scoring risks, and generating
recommendations. Agents share a persistent context that grows as the assessment
proceeds, so later agents build on what earlier ones concluded---the mechanism
that distinguishes this from standard sequential agent pipelines.

We tested it on a 15-person HIPAA-covered healthcare company and compared
outputs to independent assessments by three CISSP practitioners---the system
agreed with them 85\% of the time on severity classifications, covered 92\%
of identified risks, and finished in under 15 minutes. We then ran 30
repeated single-agent assessments across five synthetic but sector-realistic
organizational profiles in healthcare, fintech, manufacturing, retail, and
SaaS, comparing a general-purpose Mistral-7B against a domain fine-tuned
model. Both completed every run. The fine-tuned model flagged threats the
baseline could not see at all: PHI exposure in healthcare, OT/IIoT
vulnerabilities in manufacturing, platform-specific risks in retail.
The full multi-agent pipeline, however, failed every one of 30 attempts
on a Tesla T4 with its 4,096-token default context window---context
capacity, not model quality, turned out to be the binding constraint.
\end{abstract}

When a small organization needs a cybersecurity risk assessment, the realistic
options are limited. Hiring a consultant for a NIST CSF-aligned engagement
runs \$15,000 to \$50,000 and takes weeks. Building the internal expertise
costs more and takes longer. Doing nothing is free, and human factors---impulsivity, low risk
awareness, and competing priorities---consistently push small organizations
toward exactly that~\cite{hadlington2017human}.
The Verizon DBIR~\cite{verizon2024dbir} found that SMBs are targeted
nearly four times more than large organizations and account for a
disproportionate share of confirmed breaches---yet most lack even a
baseline risk assessment. There is nothing mysterious about why: the
cost of a proper assessment is not proportional to organizational size,
but the harm from a breach often is.

We are not the first people to notice this gap. NIST CSF~\cite{nist2018framework},
ISO/IEC 27005~\cite{iso27005}, and half a dozen sector-specific frameworks exist
precisely because the problem is well understood. What the frameworks do not
provide is tooling: they specify what a risk assessment should cover and how
findings should be organized, but the actual analytical work---deciding which
threats are relevant to this particular organization, determining whether the
controls they have are worth anything, translating findings into priorities
that a small team can act on---still falls on whoever is doing the assessment.

We tried the obvious thing first: hand the full organizational questionnaire
to a single model and ask it to work through the assessment start to finish.
Individual sections came out looking reasonable enough. The problem showed up
across sections---a threat likelihood rated ``High'' on page four that directly
contradicted a control gap the same model had called ``adequately mitigated''
two pages back. Recommendations at the end casually ignored budget constraints
the model had explicitly noted at the beginning. The model was capable
enough; it just could not hold that many moving pieces together in one
generation window. That realization pushed us toward the multi-agent
design we describe here. Six agents,
each with one analytical job, all reading from and writing to a shared
persistent context---so the agent writing recommendations still has access to
the intake profile from step one, not just the risk scores from step five.

This paper makes four contributions. First, a six-agent decomposition of
the NIST CSF assessment workflow, coordinated through a shared persistent
context that prevents the coherence failures we saw in both single-model
and sequential-pipeline alternatives. Second, a working implementation
deployed locally---no data leaves the organization's machine, which matters
for companies sharing sensitive infrastructure details. Third, a primary
case study validated against three CISSP practitioners. Fourth, a
cross-sector study across five organization types with 30 repeated runs,
showing that domain fine-tuning substantially improves the specificity of
threat identification.

\section{Related Work}
\label{sec:related}

\subsection{What the Frameworks Actually Give You}

NIST CSF~\cite{nist2018framework}, ISO/IEC 27005~\cite{iso27005}, and
HIPAA~\cite{hipaa1996} are normative guides, not operational ones. They specify
categories and desired outcomes but stop well short of telling a practitioner
how to decide which threats are actually plausible for a given organization,
whether a control that exists on paper is doing anything useful, or how to
translate findings into priorities a small team can act on. Freund and
Jones~\cite{freund2014measuring} document how much variability enters through
those judgment calls---different assessors, same organization, meaningfully
different findings---and their Factor Analysis of Information Risk (FAIR) tries
to impose probabilistic structure on the steps where that divergence is worst.
We use agents where they use probability distributions, but the underlying
gap---too much human judgment variability in the assessment process---is the
same one.

\subsection{LLMs in Security}

Most machine learning work in cybersecurity stays in classification
territory---intrusion detection~\cite{buczak2016survey}, malware
categorization~\cite{ye2017survey}, vulnerability prediction~%
\cite{scandariato2014predicting}---where the task is well-defined and
progress shows up in F1 scores. Strategic risk assessment is nothing like
that. You are arguing about whether a particular threat actor would
realistically target a 15-person company, whether a compensating control
actually compensates for anything, and what to tell an executive who has
ten minutes and no security background. The judgment calls pile up fast,
the data is always incomplete, and two competent practitioners will
often disagree.

Recent LLM work in security has inched toward this kind of judgment:
analyzing advisories~\cite{fang2024large}, profiling threat actors from
underground forum data~\cite{samtani2017artificial}. Brown et al.~\cite{brown2020language}
showed few-shot prompting can work on tasks with no labeled data, which is
encouraging. But writing a coherent paragraph about threat actors and
maintaining a coherent threat model across six analytical stages that depend
on each other are very different problems. The paragraph is easy. The model
is hard.

One thing our cross-sector data makes concrete: a general-purpose model will
tell you the same three things---Unauthorized Access, Data Breach, Malware
Infection---whether it is looking at a hospital handling PHI or a factory
running SCADA. It literally cannot see the difference. Fine-tuning on
cybersecurity corpora fixes that blindness, though it introduces more run-to-run
variation in the outputs (something we discuss in Section~\ref{sec:cross_sector}).

\subsection{Multi-Agent Systems and the Relevant Precedent}

Multi-agent approaches in cybersecurity have mostly been about coverage,
not reasoning: distributing detection across a
network~\cite{dasgupta2011immunity}, coordinating response across
organizational boundaries~\cite{wagner2019cyber}. Those agents are
operationally specialized---each handles a different network segment or
alert type. We decompose by \emph{reasoning role} instead.

Xi et al.~\cite{xi2023rise} survey LLM-based agent architectures and discuss
task decomposition for managing complex reasoning chains. Standard frameworks
pass outputs sequentially, with each agent seeing only its predecessor's output.
Our shared persistent context differs: it is readable and writable by every
agent in the pipeline. The Mitigation Recommendation Agent has access to the
full organizational profile from intake alongside the risk register produced
four stages later. That is the coordination mechanism that addresses the
coherence failures we observed in both single-model and sequential-pipeline
experiments.

\subsection{Explainability}

Interpretability has to be baked in from day one. If practitioners cannot
trace a risk rating back to evidence, they will override it---or worse, ignore
the whole report. So every agent in our system writes a narrative alongside
its ratings. Those narratives show up in the final report, and they exist
specifically so that the reviewing practitioner can push back on anything
that looks wrong.

\section{System Architecture}
\label{sec:architecture}

\subsection{Why Six Agents}

We did not set out to build six agents. The number came from trial and error.
Our first attempt used four, combining threat modeling and control assessment
into a single agent. The combined output was noticeably worse than when we
gave each its own agent---the reasoning styles clashed. Six is where we
landed, and while we would not claim it is the only viable decomposition,
collapsing below six degraded quality in our testing.

\subsection{Agent Roles}

\paragraph{Risk Intake Agent}
Does the upfront data gathering. Industry, size, regulatory scope, what
systems they run, where their data lives---all extracted from the
questionnaire in one pass. We decided early on that this agent should flag
anything ambiguous rather than guess. If the questionnaire says ``we work
with hospitals'' but the word HIPAA never appears, the agent writes
``HIPAA applicability: unconfirmed'' and lets a human sort it out.
That exact scenario came up in our case study.

\paragraph{Threat Modeling Agent}
Takes the organizational profile and asset inventory and builds a threat
model. Nobody needs a 200-line threat catalog, least of all a 15-person
company. The job here is narrowing down to the three or four threats that
are actually plausible for \emph{this} organization---who would realistically
target them, through what vectors, exploiting which weaknesses in the stack
they described. We prompt for plausibility, not encyclopedic coverage.

\paragraph{Control Assessment Agent}
Runs concurrently with Threat Modeling since neither needs the other's
output. Starting from the intake profile, it walks through the organization's
existing controls and asks whether they would hold up against the threats
that actually matter here. ``Firewall exists'' on a checklist means nothing.
Does the firewall block the traffic patterns we care about, or is it a
default install that has never been tuned? That kind of distinction is what
this agent tries to make. The Risk Scoring Agent later combines its output
with the threat model to build the gap analysis.

\paragraph{Risk Scoring Agent}
This is where threat modeling and control assessment converge into a risk
register---likelihood and impact for each risk, with reasoning chains
attached. Of all six agents, this one gave us the most trouble. Threat
modeling hands over evidence about attack vector categories; control
assessment hands over evidence about specific configurations. Those do not
naturally line up, and getting the scoring agent to reconcile them into
consistent ratings without losing nuance took more prompt iterations than
everything else combined.

\paragraph{Mitigation Recommendation Agent}
Writes the remediation roadmap. Our first version of this agent produced
textbook-correct recommendations that were utterly impractical for a small
company with no IT staff. The fix was simple in retrospect: before generating
any recommendations, it re-reads the organizational profile from the Risk
Intake Agent. A 15-person startup gets very different advice from a 200-person
company, even when the underlying risk findings are identical.

\paragraph{Report Synthesis Agent}
Pulls everything together into a readable report. Without this agent, the
output would be five sections in five different voices---terse threat
categories followed by audit-style control findings followed by prescriptive
recommendations. Nobody wants to read that. This agent normalizes tone,
writes the executive summary, and catches contradictions (a risk called
``high'' in one section but treated as low-priority in the recommendations,
for instance).

\subsection{Shared Persistent Context: The Coordination Mechanism}

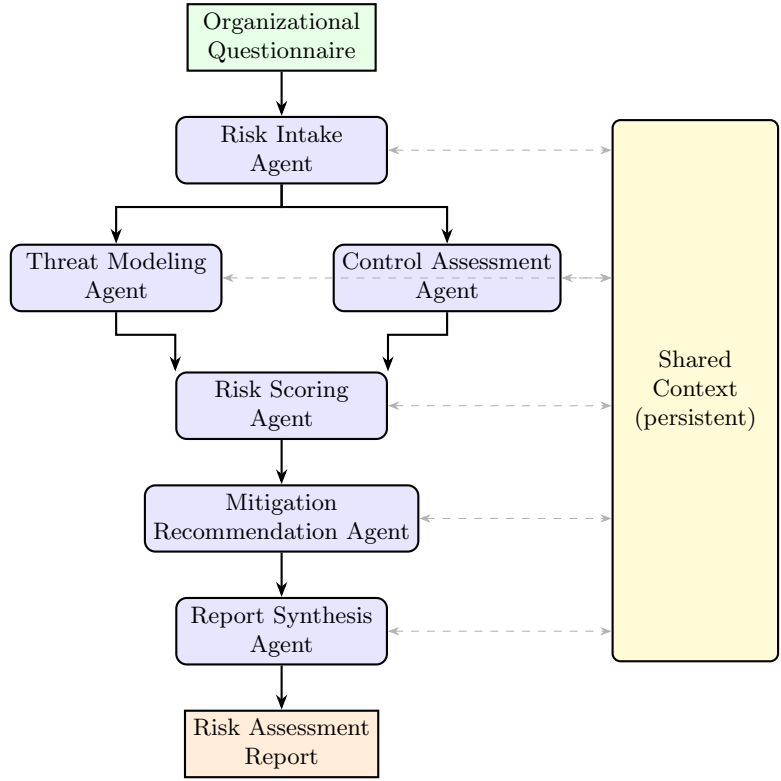
\begin{figure}[t]
\centering
\begin{tikzpicture}[
    node distance=1.1cm,
    agent/.style={rectangle, rounded corners, draw=black, fill=blue!10, thick,
      minimum width=2.8cm, minimum height=0.8cm, align=center, font=\small},
    input/.style={rectangle, draw=black, fill=green!10, thick,
      minimum width=2.5cm, minimum height=0.7cm, align=center, font=\small},
    output/.style={rectangle, draw=black, fill=orange!15, thick,
      minimum width=2.5cm, minimum height=0.7cm, align=center, font=\small},
    arrow/.style={-{Stealth[length=2mm]}, thick},
    ctx/.style={{Stealth[length=1.5mm]}-{Stealth[length=1.5mm]}, dashed,
      gray!60}
]
\node[input]  (input)      at (0, 0)    {Organizational\\Questionnaire};
\node[agent]  (intake)     at (0,-1.5)  {Risk Intake\\Agent};
\node[agent]  (threat)     at (-2.2,-3.2) {Threat Modeling\\Agent};
\node[agent]  (control)    at ( 2.2,-3.2) {Control Assessment\\Agent};
\node[agent]  (scoring)    at (0,-4.9)  {Risk Scoring\\Agent};
\node[agent]  (mitigation) at (0,-6.4)  {Mitigation\\Recommendation Agent};
\node[agent]  (report)     at (0,-7.9)  {Report Synthesis\\Agent};
\node[output] (out)        at (0,-9.4)  {Risk Assessment\\Report};

\node[rectangle, rounded corners=3pt, draw=black, fill=yellow!20, thick,
      minimum width=2.2cm, minimum height=7.2cm, align=center, font=\small]
      (kb) at (5.5,-4.7) {Shared\\Context\\(persistent)};

\draw[arrow] (input)  -- (intake);
\draw[arrow] (intake.south) -- ++(0,-0.3) -| (threat.north);
\draw[arrow] (intake.south) -- ++(0,-0.3) -| (control.north);
\draw[arrow] (threat.south)  -- ++(0,-0.3) -| (scoring.north west);
\draw[arrow] (control.south) -- ++(0,-0.3) -| (scoring.north east);
\draw[arrow] (scoring) -- (mitigation);
\draw[arrow] (mitigation) -- (report);
\draw[arrow] (report) -- (out);

\draw[ctx] (intake.east)     -- (kb.west |- intake.east);
\draw[ctx] (threat.east)     -- ++(0.5,0) |- (kb.west |- threat.east);
\draw[ctx] (control.east)    -- (kb.west |- control.east);
\draw[ctx] (scoring.east)    -- (kb.west |- scoring.east);
\draw[ctx] (mitigation.east) -- (kb.west |- mitigation.east);
\draw[ctx] (report.east)     -- (kb.west |- report.east);
\end{tikzpicture}
\caption{Six-agent pipeline with shared persistent context. Solid arrows show
sequential data handoffs; dashed bidirectional arrows show read-write access
to the shared context.}
\label{fig:architecture}
\end{figure}

Figure~\ref{fig:architecture} shows the full pipeline. The dashed lines are
the important part. In a standard pipeline, agent five only sees what agent
four handed it. By that point, the organizational profile from step one is
gone. Our early prototypes had this exact problem: the recommendation agent
would suggest things the organization clearly could not do, because it had
no way to check against the intake profile. The shared context---every agent
reads it, every agent writes to it---is what makes the later stages aware of
everything the earlier stages learned. Threat Modeling and Control Assessment
run in parallel off the Risk Intake output; the Risk Scoring Agent merges
their findings.

The pipeline is implemented in Python. Each agent is a discrete function
that receives a structured JSON payload, invokes the local LLM via the
Ollama API, and writes its output back to the shared context as a typed
JSON object. Passing structured JSON rather than raw prose means downstream
agents can reliably extract specific fields---risk scores, control gaps,
asset categories---without parsing free text.

\section{Implementation}
\label{sec:implementation}

\subsection{Model Selection and Local Deployment}

The production pipeline uses
\texttt{saki007ster/\allowbreak CybersecurityRiskAnalyst}, a
cybersecurity domain fine-tuned model deployed locally via Ollama.
For the primary case study the model ran on an RTX~4090 with a 128K
context window; the cross-sector ablation (Section~\ref{sec:model_config})
used the same model on a Tesla T4. The model was trained on security
frameworks, threat intelligence, and risk assessment corpora.
Local deployment was non-negotiable: organizations sharing
detailed infrastructure information---asset inventories, documented
control gaps, incident history---with an external API are creating a
security exposure on top of the one they are trying to assess.

The backend is a Python orchestration layer that handles agent sequencing and
context handoffs. We use Chroma as the vector database for retrieval-augmented
access to framework knowledge, and JSON schemas to enforce the input/output
contract each agent must follow. Session state lives in Supabase (PostgreSQL
underneath), and the final reports are generated from templates into PDF.

\subsection{Model Configuration Comparison}
\label{sec:model_config}

To isolate domain fine-tuning's contribution to threat identification quality,
we ran an ablation study comparing two model configurations in single-agent
mode across five organization types:

\begin{itemize}
\item \textbf{Mistral-7B-Instruct-v0.3}
  (\texttt{mistral:\allowbreak latest}): a general-purpose 7B
  instruction-tuned model with no cybersecurity-specific fine-tuning,
  serving as the baseline.
\item \textbf{saki007ster/\allowbreak CybersecurityRiskAnalyst}
  (\texttt{latest}): a cybersecurity domain fine-tuned model trained on
  security frameworks, threat intelligence, and risk assessment corpora.
\end{itemize}

Both models were deployed locally via Ollama on a Tesla T4 GPU (15\,GiB VRAM).
Each configuration was run three times per organization for repeatability,
yielding 30 total runs per model (five organizations, three runs each).

\subsection{Prompt Engineering}

We did not arrive at role-specific prompts through principled design---we
arrived through failure. The first version gave every agent the same
general-purpose system prompt, changing only the task description. Risk Intake
and Report Synthesis worked fine with this, since they are mostly about
organizing data. Control Assessment was a different story. The model kept
producing compliance assessments that mentioned controls by category
(``access management,'' ``logging'') without ever citing a specific framework
identifier. We had to add explicit grounding instructions before it would
anchor claims to indexed NIST CSF or CIS content. Threat Modeling had a
different problem: it defaulted to generating an encyclopedic threat catalog
until we added instructions forcing it to reason about the specific
organization's attack surface. Schema validation catches what still slips
through---structurally malformed JSON that would break downstream agents.

\subsection{Framework Citation Hallucinations}

This was the reliability problem we spent the most time on. The model would
confidently cite NIST CSF control identifiers and CIS Benchmark numbers that
simply did not exist, and one bad citation is enough to make a practitioner
distrust the entire report. Our fix has two parts. First, before each agent
runs, we retrieve the relevant framework text from the knowledge base and tell
the agent to reference only what it has been given---not what it ``remembers.''
Second, every framework citation in the output gets checked against our indexed
content; anything without a match gets flagged for human review instead of
making it into the report. During the full case study, this mechanism caught
seven fabricated control identifiers. Not a huge number, but enough to matter.

\section{Evaluation}
\label{sec:evaluation}

\subsection{Primary Case Study Setup}

The target organization is a U.S.-based health data analytics company with 15
employees. They process HIPAA-covered data under contract with a regional
hospital network, have no one on staff whose job title includes the word
``security,'' and rated their own maturity at 2 out of 10. They had never
gotten a formal risk assessment done---partly because they did not know what
one would cost, partly because they did not know who to call. We also
recruited three CISSP-certified practitioners who independently assessed
the same company using their standard methodology (roughly two days of work
each).

\subsection{Risk Identification}

Table~\ref{tab:risks} shows the seven high-priority risks the system
identified, with likelihood and impact ratings from the Risk Scoring Agent.

\begin{table}[h]
\caption{Top Risks Identified for Health Analytics Case Study}
\label{tab:risks}
\begin{center}
\begin{tabular}{lcc}
\toprule
\textbf{Risk} & \textbf{Likelihood} & \textbf{Impact} \\
\midrule
Lack of Security Policies              & High   & High   \\
Insufficient Authentication Controls  & High   & Medium \\
Inadequate Incident Response Plan      & High   & High   \\
Data Security \& Privacy              & High   & High   \\
Unsecured Firewall Configuration       & Medium & High   \\
Insufficient Cloud Security Controls   & Medium & Medium \\
Third-Party \& Supply Chain Security   & High   & Medium \\
\bottomrule
\end{tabular}
\end{center}
\end{table}

All three practitioners flagged the top four. Cloud security and third-party
risks were flagged by two of the three; the firewall configuration issue by
one (who had done additional research on the specific vendor they used).
To measure severity agreement, we checked every (risk, practitioner) pair
for an exact match on the High/Medium/Low label the system assigned versus
the practitioner's independent call---18 out of 21 matched, or 85\%.
All three mismatches landed in the medium-severity bucket, which is
also where the human assessors disagreed with each other most.

For coverage, we pooled every distinct risk item any of the three
practitioners flagged ($N = 13$) and asked how many the system also caught.
It got 12 of 13, or 92\%. The one it missed is telling: one practitioner flagged a server room physical access problem she
found during an on-site visit, and another caught a specific misconfiguration
through hands-on technical testing. Those are real risks. But no questionnaire
is going to surface them---you have to physically be there or directly test
the systems. One gap worth flagging: we never computed formal inter-rater
reliability (Fleiss' $\kappa$ or similar) among the three practitioners
themselves. Without knowing how much they agreed with \emph{each other},
the 85\% number is hard to calibrate---treat it as directional, not
precise.

\subsection{Compliance and Remediation}

On NIST CSF compliance, the Control Assessment Agent's findings tracked what
the practitioners found. The company was Not Compliant on Identify (no asset
inventory, no documented risk process) and Detect (zero monitoring---no log
aggregation of any kind). Protect came back Partially Compliant because they
had basic access controls but no MFA, not even on admin accounts. Respond and
Recover were both Not Compliant: no incident response plan existed, nobody
was designated to call if something went wrong, and backups had never been
tested.

The remediation roadmap the Mitigation Recommendation Agent generated was
phased across 90 days: get a security policy written first (\$3K--\$6K for a
consultant), then roll out MFA on all accounts (\$0--\$500/year in tooling),
then build an incident response plan with assigned roles by day 90. Longer-term
items---EDR deployment at \$10K--\$20K and a cloud security audit at
\$5K--\$10K---were flagged as important but outside the initial window. The
three practitioners independently rated the generated report on a 1--5
Likert scale (1\,=\,Poor, 5\,=\,Excellent); mean scores were 4.2/5 on
clarity and 4.0/5 on actionability.
Their biggest complaint was fair: the timeline was ``optimistic for a company
with no IT staff and no security budget.'' The agent does not yet know the
difference between a company that can execute changes tomorrow and one that
needs to hire someone first.

\subsection{Runtime}

The full assessment ran in 14 minutes 38 seconds on an NVIDIA RTX 4090. Human
assessments were approximately 16 person-hours each. The more important
comparison is not speed but access: 16 person-hours of CISSP-level consulting
time is simply not available to most small organizations.

\subsection{Cross-Sector Generalizability Study}
\label{sec:cross_sector}

One company is not enough to claim generalizability, and one model
configuration tells you nothing about what the fine-tuning actually
contributes. So we ran a broader study: five anonymized representative cases across
different sectors (health\_15, fintech\_30, mfg\_40, retail\_20, saas\_25---the
numbers are employee counts; these are not named real organizations but
structured anonymized profiles constructed to reflect realistic sector
characteristics), both model configurations, single-agent mode, three
runs each. That gives us 30 completed runs per model to work with.

\paragraph{Structural Reliability.}
Every single run---all 30 of them, both models---completed successfully and
produced the expected output structure (three threats, three risk statements,
three control recommendations). No structural failures, no malformed JSON, no
partial outputs. Structural stability was 1.0 across the board. The Mistral-7B
runs averaged 70.4 seconds each; the fine-tuned model was marginally slower at
72.1 seconds. Both on the Tesla T4.

\paragraph{Domain Specificity.}
Table~\ref{tab:domain_specificity} contrasts threat identifications from each
model for the same five organizations. The headline finding: Mistral-7B
produced functionally identical threat categories (Unauthorized Access, Data
Breach, and a third generic category) for all five organizations regardless of
sector or technology stack. The fine-tuned model identified threats grounded in
each organization's actual context.

\begin{table}[h]
\caption{Domain Specificity: Top Threats Identified per Sector (First Run)}
\label{tab:domain_specificity}
\begin{center}
\begin{tabular}{>{\raggedright\arraybackslash}p{1.5cm}
                >{\raggedright\arraybackslash}p{4.3cm}
                >{\raggedright\arraybackslash}p{4.3cm}}
\toprule
\textbf{Sector} & \textbf{Mistral-7B (baseline)} & \textbf{FT-CyberSec (domain)} \\
\midrule
Healthcare  & Unauthorized Access, Data Breach, Malware Infection
            & \textbf{Unsecured PHI}, \textbf{Unpatched FHIR Integrations},
              Insufficient Authentication \\
\addlinespace
Fintech     & Unauthorized Access, Data Breach, Insecure Third-Party Vendors
            & Unauthorized Access to PII, Data Breach (Insufficient Controls),
              Denial of Service \\
\addlinespace
Mfg.        & Unauthorized Access, Data Breach, Malware Infection
            & \textbf{Unpatched Windows AD}, \textbf{OT/IIoT Sensor Vulnerabilities},
              Insufficient Segmentation \\
\addlinespace
Retail      & Unauthorized Access, Data Breach, Phishing Attacks
            & \textbf{Unpatched Vulns in Shopify}, \textbf{Insufficient Controls
              for AWS}, Phishing Attacks \\
\addlinespace
SaaS        & Unauthorized Access, Data Leakage, Insecure Code
            & Unauthorized Access to PII, Data Breach (Insufficient Measures),
              DoS Attack \\
\bottomrule
\end{tabular}
\end{center}
\end{table}

Look at the healthcare row: the fine-tuned model flags unsecured PHI and
unpatched FHIR integrations---the baseline says ``Unauthorized Access'' and
``Data Breach,'' which could apply to literally any organization on earth. The
manufacturing row is equally stark: OT/IIoT sensor vulnerabilities versus
generic ``Malware Infection.'' A practitioner handed the baseline output would
learn nothing they did not already know.

\paragraph{Output Variability.}
Structure stayed locked across all three runs per organization for both models.
The interesting difference is in what the models actually said. Mistral-7B
repeated itself heavily---3 to 4 unique threat titles across three runs,
sometimes completely identical across all three. The fine-tuned model produced
6 to 9 unique titles across three runs on the same organization. That is the
price of specificity: a model sampling from a richer, more context-aware
distribution will give you different answers each time, even for the same
input. We think that is a feature, not a bug---but it does mean a single run
should not be taken as the final word.

\paragraph{Multi-Agent Pipeline Completion.}
The multi-agent pipeline achieved 0\% completion across all 30 runs in this
environment, compared to 100\% for single-agent runs.
Section~\ref{sec:scalability} discusses this finding.

\section{Discussion}
\label{sec:discussion}

\subsection{Why the Decomposition Matters More Than We Expected}

The multi-agent split was supposed to buy us specialization. It did, but that
was not the main thing. Coherence was. A single model writing a full
assessment produces sections that look fine individually, but the report drifts
as a whole. Page eight contradicts page three. The recommendation section
ignores constraints acknowledged in the introduction. Practitioners spot these
inconsistencies fast, and once they do, the whole report goes in the bin.

Splitting into six agents helped, but not as much as the shared context did.
In our early pipeline, agents passed their output forward to the next one and
that was it. By the time the recommendation agent ran, it had no way to look
back at the intake profile from five steps ago. The fix was straightforward:
give every agent read-write access to a single persistent context object.
That one change did more for output quality than anything else we tried.
We should be upfront that the evidence here is developmental---we watched
it happen across iterations, but we did not run a controlled ablation
pitting shared-context against sequential-pipeline against single-model
on identical inputs. That experiment would nail the point down; for now
the claim rests on what we observed while building the system.

\subsection{Domain Fine-Tuning and the Specificity--Consistency Tradeoff}

The cross-sector data makes the fine-tuning argument concrete. Mistral-7B told
every organization the same thing: your top threats are Unauthorized Access and
Data Breach. Hospitals, factories, retail---did not matter. The fine-tuned model
actually looked at the input. It told the hospital about PHI exposure and
unpatched FHIR integrations. It told the factory about OT/IIoT sensor
vulnerabilities. Those are actionable findings. ``Unauthorized Access'' is not.

There is a cost, though, and we should not gloss over it. Run the fine-tuned
model three times on the same company and you will get 6 to 9 different threat
titles. Mistral-7B gives you 3 or 4, and sometimes all three runs are word-for-word
identical. Generic answers are trivially reproducible precisely because they
ignore context. The fine-tuned model's outputs vary because it is actually
responding to what it was given. We think that is the right tradeoff, but it
means any single run should be treated as one sample, not the final answer.
A caveat: these cross-sector profiles are synthetic, so we cannot be sure
whether the fine-tuned model is actually reasoning about threats or just
pattern-matching on sector keywords in the input. Testing on real
companies across industries would settle that question; for now, it
remains open.

\subsection{Where It Breaks Down}

We want to be direct about where the system falls short.

The most dangerous failure mode is that it writes convincingly even when fed
bad data. An organization that underreports incidents or overstates its
controls will get an optimistic-sounding assessment. Human assessors push back
on inconsistent answers. Our agents do not.

The system also cannot see anything the questionnaire does not mention. It
will never find the server room door that does not lock or the misconfigured
firewall rule---those require physical presence or hands-on testing. That is
where the 8\% coverage gap in the primary case study came from.

Third, the model knows only what it was trained on. Threat techniques from
after the fine-tuning cutoff get less weight than a current practitioner would
give them. And fourth, the output variability we documented means you should
run the fine-tuned model more than once before treating any result as settled.

\subsection{Multi-Agent Pipeline Scalability}
\label{sec:scalability}

None of the multi-agent runs completed on the Tesla T4. Zero out of thirty.
Meanwhile, every single-agent run finished without issue. The primary case
study on an RTX 4090 (128K context window) had not shown this problem at all,
so it caught us off guard. The root cause is prosaic: the Tesla T4 defaults to
a 4,096-token context window, and the six-agent pipeline---which passes growing
structured JSON through the shared context---blows past that limit by the third
or fourth agent. Single-agent runs need only one model call, so they fit fine.

This points to a structural requirement that anyone deploying multi-agent
pipelines should plan for explicitly: multi-agent architectures are
categorically more demanding on context window size and VRAM than
single-agent equivalents. The demand scales with the number of agents and
the size of the shared context, not just the individual model's footprint.
A pipeline that works cleanly on a high-memory workstation will fail silently
on constrained hardware---not because the model is wrong, but because the
context accumulation outpaces available capacity. In our case, 128K context
and 24\,GiB VRAM on the RTX~4090 was sufficient; 4,096 tokens and 15\,GiB
on the T4 was not. Practitioners targeting deployment on cloud GPUs or
edge hardware should treat context window capacity as a first-class
infrastructure constraint.

\subsection{A Note on Adversarial Use}

We have not evaluated the system's own attack surface, and we should say so.
An insider who wants to hide a control gap---or an external attacker who has
compromised the questionnaire pipeline---could feed the agents misleading
data in ways that a human assessor would likely catch but the current system
would not. That risk needs to be addressed before production use.

\section{Conclusion}
\label{sec:conclusion}

We built a six-agent system for automated cybersecurity risk assessment. On
a real 15-person company, it agreed with three CISSP practitioners 85\% of
the time on severity and covered 92\% of their findings, finishing in about
15 minutes. A broader study across five sectors and 30 runs per model
confirmed that the fine-tuned model catches sector-specific threats the
baseline cannot see. The multi-agent pipeline choked on constrained GPU
hardware---zero completions out of 30 attempts on the Tesla T4---and fixing
that remains open work.

The part we think matters most: a company with a real HIPAA compliance
obligation, no security program, and no path to understanding it can get a
credible, sector-specific starting point in 15 minutes. That is not
expert-quality assessment. But it is categorically more useful than what most
small organizations have, which is nothing.

\section*{Data Availability}
Run-level data---completion rates, latency, and the threat titles each
model produced---are available from the corresponding author on request.
The five organizational profiles used in the cross-sector study are
entirely synthetic and contain no real company data.

\section*{Disclosure of Interests}
This work was conducted independently and does not relate to the authors'
positions at Microsoft, BigCommerce, or Amazon, nor does it represent the
views or interests of these organizations. The authors have no competing
interests to declare that are relevant to the content of this article.


\end{document}